\documentclass[12pt]{article}
\usepackage{graphicx} 
\newcommand{\e}{{\rm e}}

\newcommand{\maxrm}{{\rm max}}

\newcommand{\ug}{\; = \;}

\newcommand{\Vbf}{\mbox{\boldmath $V$}}

\newcommand{\kbf}{\mbox{\boldmath $k$}}
\newcommand{\imp}{\mbox{\boldmath $p$}}

\newcommand{\nablabf}{\mbox{\boldmath $\nabla$}}

\newcommand{\pa}{\partial}
\newcommand{\la}{\lambda}
\newcommand{\La}{\Lambda}

\newcommand{\text}{\rm}

\newcommand{\drm}{{\rm d}}

\newcommand{\Acal}{{\cal A}}
\newcommand{\Aove}{{\overline A}}
\newcommand{\Scal}{{\cal S}}
\newcommand{\Ncal}{{\cal N}}

\newcommand{\al}{\alpha}

\newcommand{\bb}{\begin{equation}}
\newcommand{\ee}{\end{equation}}
\newcommand{\bega}{\begin{eqnarray}}
\newcommand{\ega}{\end{eqnarray}}
\newcommand{\begae}{\begin{eqnarray*}}
\newcommand{\egae}{\end{eqnarray*}}

\newcommand{\h}{\hspace*{4ex}}
\newcommand{\dis}{\displaystyle}

\newcommand{\ove}{\overline}

\newcommand{\cent}{\centerline}
\newcommand{\vs}{\vspace*}

\begin{document}

\baselineskip 0.65cm

\begin{center}

{\large {\bf Soliton-like solutions to the ordinary Schroedinger
equation}$^{\: (\dag)}$} \footnotetext{$^{\:
(\dag)}$  Work partially supported by INFN and MIUR (Italy), and
by FAPESP (Brazil). \ E-mail addresses for contacts:
recami@mi.infn.it [ER]; mzamboni@ufabc.edu.br [MZR]}

\end{center}

\vs{5mm}

\cent{ Michel Zamboni-Rached, }

\vs{0.2 cm}

\centerline{{\em DMO, FEEC, UNICAMP, Campinas, SP, Brasil}}

\vs{0.2 cm}

\centerline{\rm and}

\vs{0.3 cm}

\cent{ Erasmo Recami }

\vs{0.2 cm}

\cent{{\em Facolt\`a di Ingegneria, Universit\`a statale di
Bergamo, Bergamo, Italy;}}

\cent{and {\em INFN---Sezione di Milano, Milan, Italy.}}

\vs{0.8 cm}

{\bf Abstract  \ --} \ In recent times it has been paid attention to the fact that (linear) wave
equations admit of ``soliton-like" solutions, known as Localized Waves or Non-diffracting Waves,
which propagate without distortion in one direction. Such Localized Solutions (existing also for
K-G or Dirac equations) are a priori suitable, more than gaussian's, for describing elementary
particle motion. \ In this paper we show that, {\em mutatis mutandis}, Localized Solutions exist
even for the ordinary Schroedinger equation within standard Quantum Mechanics; and we obtain both
approximate and exact solutions, also setting forth for them particular examples. \ In the ideal
case such solutions bear infinite energy, as well as plane or spherical waves: we show therefore
how to obtain finite-energy solutions. \ At last, we briefly consider solutions for a particle
moving in the presence of a potential.

\

\


\

\


\newpage

\section{Introduction}

Recently it has been shown ---as it had been already realized in
old times[1]--- that not only nonlinear, but also a large class of
linear equations (including, in particular, the wave equations)
admit of ``soliton-like" solutions. \ Those solutions[2] are {\em
localized,} and travel along their propagation axis practically
without diffracting (at least until a certain field-depth[2,3,4]):
Such wavelets were indeed called ``undistorted progressing waves"
by Courant and Hilbert[1].  Let us recall that their peak-velocity
$V$ can assume any values[5,6,2] $0 \leq V \leq \infty$, even if we
are mainly interested here in their {\em localization properties}
rather than in their group-velocity. \ In the case of wave
equations, the localized solutions more easy to be constructed in
exact form resulted to be the so-called  ``(superluminal) X-shaped"
ones (see Refs.[4,7,8,2], and refs. therein).

\h The X-shaped waves, long ago predicted[6] to exist within Special
Relativity (SR), have been first mathematically constructed[9,2] as solutions
to the wave equations in Acoustics[4], and later on in Electromagnetism
(namely, to the Maxwell equations[7]), and soon after produced
experimentally[10]. \ Only very recently, {\em subluminal} localized
solutions have been suitably worked out in exact form[11],
even for the case of zero speed (``Frozen Waves").[12]

\h It was soon thought that, since the mentioned solutions to the wave
equations are non-diffractive and particle-like, they may well be related
to elementary particles (and to their wave nature)[13,14]. And, in fact, localized
solutions have been found for Klein-Gordon and for Dirac equations[13,14].

\h However, little work[15] has been done, as far as we know, for the
({\em different\/}) case of the {\em Schroedinger} equation\footnote{For some
work in connection with the ordinary Schroedinger equation, see for instance,
besides [7], also Refs.[14].}. Indeed, the relation between the
energy $E$ and the
impulse magnitude $p \equiv |\imp|$ is quadratic [$E=p^2/(2m)\/$] in the
non-relativistic case, like in Schroedinger's, at variance with the
relativistic one.  But, as we were saying, the nondiffracting solutions,
which are essentially superpositions of Bessel beams and are
currently called {\em Localized Waves}, would be quite apt at
describing elementary particles: much more than the gaussian waves.
In this paper we show that indeed, {\em mutatis mutandis}, Localized Solutions exist
even for the ordinary Schroedinger equation within standard Quantum Mechanics;
and we obtain both approximate and exact solutions,
also setting forth for them particular examples. In the ideal case such solutions bear infinite
energy, as well as spherical or plane waves: we shall therefore show how to obtain finite-energy solutions. \
At last, we shall briefly consider solutions for a particle moving in the presence of a potential.

\h Before going on, let us recall that, in the time-independent
realm ---or, rather, when the dependence on time is only harmonic,
i.e., for monochromatic solutions---, the (quantum,
non-relativistic) Schroedinger equation is mathematically
identical to the (classical, relativistic) Helmholtz
equation[16]. \ And many trains of localized
X-shaped pulses have been found, as {\em superpositions} of
solutions to the Helmholtz equation, which propagate, for
instance, along cylindrical or co-axial waveguides[17]; \ but we
shall skip all the cases[18] of this type, even if interesting,
since we are concerned here with propagation in free space, even when in the
presence of an ordinary potential. \ Let
us also mention that, in the {\em general} time-{\em dependent}
case, that is, in the case of pulses, the Schroedinger and the
ordinary wave equation are no longer mathematically identical,
since the time derivative results to be of the fist order in the
former and of the second order in the latter. [It has been shown
that, nevertheless, at least in some cases[19], they still share
various classes of analogous solutions, differing only in their
spreading properties[19]]. Moreover, the
Schroedinger equation implies the existence of an {\em intrinsic}
dispersion relation even for free particles.

\h Another difference, to be kept here in mind, between the wave and the
Schroedinger equations is that the solutions to the wave equation suffer
only diffraction (and no dispersion) in the vacuum, while those of the
Schroedinger equation suffer also (an intrinsic) dispersion even in the
vacuum.

\h Let us repeat that the majority of the ideal localized solutions
we are going to construct are endowed with infinite energy. We shall treat
also a {\em finite-energy}
case\footnote{In such cases the solutions travel undistorted and
with a constant speed along a finite depth of field only.} only towards the end
of this paper: In fact, infinite-energy solutions themselves, even without
truncating them in space and time, results to be rather useful for describing
wavepackets in regions not too extended in the transverse direction; as we shall
see below.

\

\section{Bessel {\em beams} as localized solutions (LS) to the Schroedinger
equation}

Let us consider the Schroedinger equation for a free particle (an electron,
for example)

\bb
\nablabf^2 \psi + {{2im} \over \hbar} \, {{\pa \psi} \over {\pa t}}  \ug 0 \; .
\label{eq1}
\ee      %%%eq.1

If we confine ourselves to solutions of the type

\

\hfill{$
\psi(\rho,z,\varphi;t) \ug F(\rho,z,\varphi) \; \e^{-iEt/\hbar} \; ,
$\hfill}

\

their spatial part $F$ obeys the reduced equation

\bb
\nablabf^2 F + k^2 F  \ug 0 \; ,
\label{eq2}
\ee      %%%eq.2

with \ $k^2 \equiv p^2/\hbar^2$ \ and \ $p^2 = 2mE$ (quantity $p
\equiv |\imp|$ being the particle momentum, and therefore $k \equiv
|\kbf|$ the total wavenumber). \ Equation (2) is nothing but the
Helmholtz equation, for which various simple localized-beam
solutions (LS) are already known: In particular, the so-called
Bessel beams[2], which have been experimentally produced since
long[20].

\h Namely, let us now look ---as usual--- for solutions
(cylindrically symmetric with respect to [w.r.t.] the $z$-axis) of
the form

\

\hfill{$
\psi(\rho,z;t) \ug R(\rho) \ Z(z) \ T(t) \; ,
$\hfill}

\

and explicitly indicate, mainly for clarity's sake, the subsequent steps. \
Equation (\ref{eq1}) then becomes

\bb
\left[ {1 \over {\rho R}} {\pa \over {\pa\rho}} \left( \rho {{\pa R}
\over {\pa\rho}} \right) + {1 \over z} {{\pa^2 Z} \over {\pa z^2}} \right]
\ug -i {{2m} \over \hbar} {1 \over T} {{\pa T} \over {\pa t}}
\label{eq3}
\ee      %%%eq.3

so that

\bb
-{{i2m} \over {\hbar T}} {{\pa T} \over {\pa t}} = -k^2 \ \Longrightarrow \
T = \e^{-iEt/\hbar}
\label{eq4}
\ee      %%%eq.4

\

where, let us repeat, $E = p^2/(2m) = k^2 \hbar^2/(2m)$ (and in fact the
last exponential is often written as $\exp{[-i\omega t]}$).

\h Analogously, we have

\bb
{1 \over {\rho R}} {\pa \over {\pa\rho}} \left( \rho {{\pa R} \over
{\pa\rho}} \right) + k^2 \ug - {1 \over z} {{\pa^2 z} \over {\pa z^2}}
\label{eq5}
\ee      %%%eq.5

and therefore

\bb - {1 \over z} {{\pa^2 z} \over {\pa z^2}} \ug k_z^2 \
\Longrightarrow \ \dis{z = \e^{izk_z}} \; , \label{eq6}
\ee      %%%eq.6

\

where the constant $k_z \equiv k_\parallel = p_\parallel/\hbar
\equiv p_z/\hbar$ is the longitudinal wavenumber.\footnote{Since
the present formalism is used both in quantum mechanics and in
electromagnetism, with a difference in the customary nomenclature,
for clarity's sake let us here stress, or repeat, that $k \equiv
p/\hbar$; $k_\rho \equiv k_\perp \equiv p_\perp/\hbar$; $\omega
\equiv E/\hbar$; while $k_z \equiv k_\parallel = p_\parallel
/\hbar \equiv p_z/\hbar$ is often represented by the (for us)
ambiguous symbol $\beta$.} \ We will suppose $k_z \geq 0$, that
is, $p_z \geq 0$, to ensure that we deal with forward traveling
beams only.

\h As a consequence, the (transverse) function $R = R(\rho)$ obeys
the equation

\bb
{1 \over {\rho}} {\pa \over {\pa\rho}} \left( \rho {{\pa R} \over {\pa\rho}}
\right) + (k^2 - k_z^2) R \ug 0
\label{eq7}
\ee      %%%eq.7

which is a Bessel differential equation admitting as solution the Bessel
function\footnote{The other Bessel functions are not acceptable here, because
of their divergence at $\rho = 0$ or for $\rho \rightarrow \infty$.}

\bb
R(\rho) \ug J_0(\rho k_\rho) \; ,
\label{eq8}
\ee      %%%eq.8

where the constant $k_\rho \equiv k_\perp \equiv p_\perp /\hbar$ is the
transverse wavenumber, and

\bb
k_\rho^2 = k^2 - k_z^2 \equiv 2mE/{\hbar^2} - k_z^2 \; .
\label{eq9}
\ee      %%%eq.9

\begin{figure}[!h]
\begin{center}
 \scalebox{.5}{\includegraphics{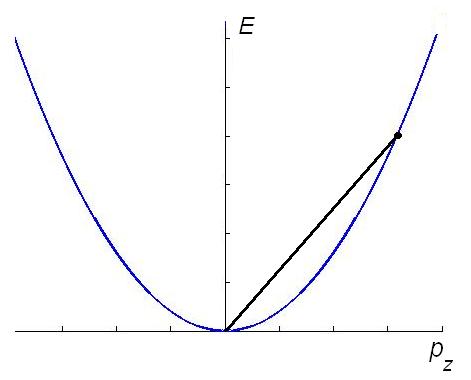}}
\end{center}
\caption{The parabola and the chosen straight-line have
equations $E = p_z^2/(2m)$ and $E = Vp_z$, respectively. The
intersection of our straight-line with the parabola corresponds to
the value $E = 2mV^2$. \ The allowed region is the one internal to the
parabola, since it must be $E \geq p_z^2/(2m)$.} \label{fig1}
\end{figure}

To avoid any divergencies, it must be $k_\rho^2 \geq 0$, that is,
$k^2 \geq k_z^2$; namely, it must hold [see (a) in Fig.\ref{fig1}]
the constraint

\bb
E \, \geq \, {p_z^2 \over {2m}}
\label{eq10}
\ee      %%%eq.10

[Notice that, to avoid the appearance of evanescent waves, one
should postulate $k_z$ to be real; but such a condition is already
included in our previous assumption that $k_z \geq 0$]. \ In the
following, to simplify the notations, we shall also put [$p \equiv \hbar k \,$]:

$$p_\rho \, \equiv \, p_\perp \ .$$

\h The solution is therefore:

\bb
\psi(\rho,z;t) \ug J_0(\rho p_\rho/\hbar) \; \exp{[i(z p_z - Et)/\hbar]}
\label{eq11}
\ee      %%%eq.11

together with condition (9). \ {\em Equation {\rm (11)} can be regarded
as a Bessel beam solution to the Schroedinger equation.}  \ This result
is not surprising, since ---once we suppose the whole time variation to be
expressed by the function $\exp{[i\omega t]}$--- both the
ordinary wave equation and the Schroedinger equation transform
into the Helmholtz equation. \ Actually, the only difference
between the Bessel beam solutions to the wave and to the
Schroedinger equation consists in the different relationships
among frequency, longitudinal, and transverse wavenumber;
in other words (with $E \equiv \omega\hbar$):

\

\hfill{$ p_\rho^2 \ug E^2/c^2 - p_z^2 \ \ \ \ \ {\rm for \
the \ wave \ equation};
$\hfill} (12a)          %%%eq.(12a)

\hfill{$ p_\rho^2 \ug 2mE - p_z^2 \ \ \ \ \ \ {\rm for \ the \
Schroedinger \ equation}. \
$\hfill} \ \ (12b)          %%%eq.(12b)
\setcounter{equation}{12}

\

\h In the case of beams, the experimental production of LSs to the
Schroedinger equation can be similar to the one exploited for the
LSs to the wave equations (e.g., in Optics, or Acoustics):  Cf.,
e.g. Figure 1.2 in the first one of Refs.[8],
and refs. therein, where the simple
case of a source consisting in an array of circular slits, or
rings, were considered.\footnote{For pulses, however, the
generation technique must deviate from Optics', since in the
Schroedinger equation case the phase of the Bessel beams produced
through an annular slit would depend on the energy.}  \ In the
Table we refer to a Bessel beam of photons, and a Bessel beam of
(e.g.) electrons, respectively. We list therein the relevant
quantities having a role, e.g., in Electromagnetism, and the
corresponding ones for the Schroedinger equation's spatial part
$\hbar^2 \nablabf^2 F + 2mE \, F = 0$, with $F = R(\rho) \;
Z(z)\,$. The second and the fourth lines have been written down
for the simple Durnin et al.'s case, when the Bessel beam is produced
by an annular slit (illuminated by a plane wave) located in the focus
of a lens[20].

\

\

\begin{tabular}{|c|c|}
  \hline
  WAVE EQUATION & SCHROEDINGER EQUATION \\
  \hline
  $k  \ug {\omega \over c}$ & $p \ug \sqrt{2mE}$ \\
  $k_\rho \, \simeq \, {r \over f} \, k$ & $p_\rho \simeq {r \over f} \, p$ \\
  $k_\rho^2 \ug {\omega^2 \over c^2} - k_z^2$ & $p_\rho^2 \ug 2mE - p_z^2$ \\
  $k_z^2 \ug {\omega^2 \over c^2}(1-{r^2 \over f^2})$ & $p_z^2 \ug 2mE (1-{r^2 \over f^2})$ \\
  \hline
\end{tabular}

\

\

In this Table, quantity $f$ is the focal distance of the lens (for
instance, an ordinary lens in optics; and a magnetic lens in the case of
Schroedinger charged wavepackets), and $r$ is the radius of the
considered ring. [In connection with the last line of the Table, let us recall
that in the wave equation case the phase-velocity $\omega / k_z$
is almost independent of the frequency (at least for limited frequency
intervals, like in optics), and one gets a constant group-velocity and
an easy way to build up X-shaped waves. By contrast,
in the Schroedinger case, the phase-velocity of each (monochromatic)
Bessel beam depends on the frequency, and this makes it difficult to
generate an ``X-wave" (i.e., a wave depending on $z$ and $t$ only via the
quantity $z-Vt$) by using simple methods, as Durnin et al.'s,
based on Bessel beams superposition. In the case of charged particles,
one should compensate such a velocity variation by suitably modifying
the focal distance $f$ of the Durnin's lens, e.g. on having recourse
to an additional magnetic, or electric, lens.]

\h Before going on, let us stress that one could easily eliminate
the restriction of axial symmetry: In such a case, in fact,
solution (11) would become

\

\hfill{$
\psi(\rho,z,\varphi;t) \ug J_n(\rho p_\rho/\hbar) \ \dis{\e^{i z p_z/\hbar} \, \e^{-iEt/\hbar}
\, \e^{i n \varphi}} \; ,
$\hfill}

\

with $n$ an integer. \ The investigation of not
cylindrically-symmetric solutions is interesting especially in the
case of localized {\em pulses} (cf. Sect.3): and we shall deal
with them below.

\

\section{Localized {\em pulses} as solutions to the
Schroedinger equation ({\em approximate method\/})}

Localized (non-dispersive, besides non-diffracting) {\em pulses}
can be constructed, as solutions to the Schroedinger equation, both by
having recourse to the standard ``paraxial approximation", and in an
exact, analytic way. \ Let us start with the approximate method.

\h Let us go back, then, to our Bessel beam solution (11), with condition
(10).  \ We can obtain localized (non-dispersive) pulses, as
solutions to Schroedinger's equation, by suitably superposing the beam
solutions (11), and by selecting in the plane $(p_z,E)$ the straight-line
[see Fig.1]:

\bb
E \ug V p_z \; ; \ \ \ \ \ \ \ \ \ \ \ \ \ \ \ \ \ (p_z \geq 0) \ ,
\label{eq13}
\ee      %%%eq.13

vith $V$ a chosen constant speed; so that from eq.(10) one gets the important condition

\bb
E \; \leq \; 2 m V^2
\label{eq14}
\ee      %%%eq.14

and eq.(11) can consequently be written

\

\hfill{$
\psi(\rho,\zeta) \ug J_0(\rho p_\rho /\hbar) \; \exp{[i p_z \zeta / \hbar]}
$\hfill} (11')   %%%eq.(11')

\

where now \ $p_\rho^2 = (2mE-p_z^2) = E (2m -E/V^2)$ \ and we introduced the
new variable

\bb
\zeta \; \equiv \; z - V t \; .
\label{eq15}
\ee      %%%eq.15

\h Localized-wave solutions can be therefore obtained through the superposition
(see Fig.1):

\bb \Psi(\rho,\zeta) \ug \Ncal \int_{0}^{2mV^2} \drm E \; J_0 \left( \rho
\sqrt{{E \over \hbar^2} (2m - {E \over V^2})} \right) \;
\exp{[{i {E \over {\hbar V}} \zeta}]} \; S(E)
\label{eq16}
\ee      %%%eq.16

the weight-function $S(E)$ being a suitable energy-spectrum (with the dimensions,
as usual, of the inverse of an Energy), while $\Ncal$ is a ``normalization" constant which
normalizes to 1 the peak-value of $|\Psi|^2$ and (since it
multiplies a dimensionless integral) bears the dimensions $[\Ncal]
= [L^{3/2}]$, to respect the ordinary meaning of
${|\Psi(\rho,\zeta)|}^2$. \ It should be noted that we are
integrating, in the space ($p_z,E$) along the straight-line (13),
that is, $E = V p_z \ $. \ This corresponds to superposing Bessel
beams all endowed with the same phase-velocity $V_{\rm ph} \equiv
V$. \ The resulting pulse will possess $V$ as its {\em
group-velocity} (namely, as its peak-velocity), since it is
well-known that, when the phase-velocity $V_{\rm ph}$ does not
depend on the energy or frequency, the resulting pulse happens to
travel with the group-velocity $V_{\rm g} \equiv \pa\omega/\pa k_z
= V_{\rm ph} \equiv V$: cf. refs.[17,2,21] and refs.
therein. \ Due to constraint (14), we are actually integrating
along our straight-line from $0$ to $2mV^2$ (see Fig.1).

\h It is important also to note explicitly that each solution
$\Psi(\rho,\zeta)$ given by eq.(16), depending on $z$ (and $t$)
only via the variable $\zeta \equiv z-Vt$, does represent a pulse
that appear with a constant shape to an observer traveling with
speed $V$ along the wave motion-line $z$: \ in other words, it
represents a pulse which propagates rigidly along $z$. \ {\em
Therefore, eqs.(16) are already ---as desired-- non-dispersing and
non-diffracting ("localized") solutions to the Schroedinger
equation.}

\h Integrals (16), however, appear difficult to be analytically performed,
independently of the spectrum $S(E)$ chosen.

\h To overcome this difficulty, let us rewrite eq.(11') as a function of
$p_\rho$ only, by exploiting eq.(12b), which can be written \
$E^2/V^2 - 2mE + p_\rho^2 = 0$, \ and yields

\bb E \ug mV^2 \left( 1 + \sqrt{1 - {p_\rho^2 \over p_{\rho \maxrm}^2}} \right)
\; ,
\label{eq17}
\ee      %%%eq.17

where

$$  p_{\rho \maxrm} \ug mV \; ,$$

as it comes by deriving eq.(12b) with respect to $E$.

\

Therefore, eq.(11') becomes

\

\hfill{$ \psi(\rho,\zeta) \ug  J_0(\rho p_\rho/\hbar) \, \exp{
[i{{mV} \over \hbar} \, \zeta \,  \sqrt{1 - {p_\rho^2 \over
{m^2V^2}}}\; ] } \ S(p_\rho/\hbar) \ \dis{\e^{i{{mV} \over \hbar} \zeta}}
$\hfill} (11'')   %%%eq.(11'')

\

with $0 \leq p_\rho \leq p_{\rho \maxrm}$, where,\footnote{For the sake of
clarity, let us repeat that, when the phase-velocity $V$ becomes
(as in our case) the group-velocity, $V_{\rm g} = V$, then the
component $p_\rho$ of $\imp$ acquires $mV$ as its maximum value. \
It holds, moreover, $\sqrt{p^2 -
p_\rho^2} = p_\parallel \equiv p_z$, which just equals $p$, since
in the present case $V \equiv |\Vbf| = V_z$.} let us repeat,
$p_{\rho \maxrm} = mV$. \ Then, the Localized Solutions will be
written as

\bb \Psi(\rho,\zeta) \ug \Ncal \e^{imV \zeta/\hbar} \int_{0}^{mV} \drm
p_\rho \; J_0({{\rho p_\rho} \over \hbar}) \; S(p_\rho) \; \exp{\left[ {{i mV} \over \hbar} \,
\zeta \, \sqrt{1 - {p_\rho^2 \over {m^2V^2}}} \right] } \; .
\label{eq18}
\ee      %%%eq.18

Let us notice that, in the new variable $p_\rho$, the Bessel function,
previously written as in eq.(16), gets, as we have seen, the simplified
expression $J_0(\rho p_\rho)$.

\h It is now enough to choose a weight-function $S$ that is strongly
bumped around the value $p_\rho$, in the interval [$0,mV$], with

\bb
p_\rho \ll mV \; ,
\label{eq19}
\ee      %%%eq.19

for being able to integrate from $0$ to $\infty$ with a negligible
error. \ Namely, let us now adopt the so-called {\em paraxial
approximation}. \ Under condition (19), one can approximate the
exponential factor as follows:

\

\hfill{$ mV \dis{ \sqrt{1-{p_\rho^2 \over {m^2V^2}}} \, \simeq \,
mV - {1 \over 2} {p_\rho^2 \over {mV}} } \; , $\hfill}

\

so that eq.(18) can be eventually written in terms of an integration from
$0$ to $\infty$:

\bb \Psi(\rho,\zeta) \ug \Ncal \e^{2imV \zeta/\hbar}
\int_{0}^{\infty} \drm p_\rho \; J_0({{\rho p_\rho} \over \hbar})  \ S(p_\rho) \
\exp{[i{{p_\rho^2} \over {2 \hbar mV}} \zeta]} \; .
\label{eq20}
\ee      %%%eq.20

\h Let us now examine various special cases of weight-functions $S(p_\rho)$
obeying the previous conditions: that is, well localized around a value
$p_\rho \ll mV$.

\

\subsection{Some examples of approximate Localized Solutions to the
Schroedinger equation ({\em paraxial approximation\/})}

As already claimed, we are for the moment adopting the {\em paraxial approximation,}
since it yields good, and interesting enough, results: Only
in the subsequent Sections we shall go on to the exact, analytical
approach.

\

\h {\em First of all,} let us consider the simple spectrum

\bb  S(p_\rho) \ug 4 q \, p_\rho \, \dis{\e^{-q p_\rho^2}}
\label{eq21}
\ee      %%%eq.21

(with the dimensions, now, of the inverse of an Impulse), with

\

\hfill{$ q \equiv \dis{ {\al \over {m^2 V^2}}}
$\hfill} (22a)          %%%eq.(22a)

\

so that the above conditions merely imply the dimensionless constant $a$ to be

\

\hfill{$ \al \gg 1 \; . \
$\hfill} (22b)          %%%eq.(22b)
\setcounter{equation}{22}

\

In this case, also the total spectral-width $\Delta p_\rho$ results to be
$\Delta p_\rho \ll mV$: and this too supports the fact that our integral can
indeed run till $\infty$. \ In eq.(20), one can then perform
(analytically) the integration, and get the solutions

\bb \Psi(\rho,\zeta) \, \simeq \, \Ncal \ 4 q \hbar^2 \; \e^{2imV \zeta/\hbar} \; {1 \over
{2Q}} \; \exp{[-{\rho^2 \over {4\hbar(q\hbar-i{1 \over {mV}}
\zeta)}}]} \; ,
\label{eq23}
\ee      %%%eq.23

\

quantity $q$ being still the one defined in eq.(22a), with $\al \gg 1$; while
function $Q$ is

\bb
Q \, \equiv \, \hbar (q\hbar - {i \over {2mV}} \, \zeta) \; .
\label{eq24}
\ee      %%%eq.24

Equation (23) constitutes an interesting solution of the
Schroedinger equation: It describes a wavepacket rigidly moving
with the chosen speed $V$. \ The maximum of its intensity $|\Psi|^2$
occurs at

$$\rho \ug 0 ; \ \ \ \ \ \zeta \ug 0 \; ,$$

and therefore also such a maximum travels with the speed $V$, as expected
(since $\zeta = z-Vt$). \ For $\zeta=0$ one gets [$\al \gg 1$]:

\bb  |\Psi(\rho,\zeta=0)|^2 \, \simeq \, {\Ncal}^2 \; 4 \;
\exp{[-{{\rho^2} \over {2q \hbar^2}}]} \; ,
\label{eq25}
\ee      %%%eq.25

and the {\em transverse} localization $\Delta \rho$ of the wavepacket results
to be

\

\hfill{$ \Delta \rho \ug \dis{ {\hbar \over {mV}} } \, \sqrt{2\al}
\; ,
$\hfill} (25')          %%%eq.(25')

\

which shows also the r\^ole of $\al$ (and therefore of $q$) in regulating the
wavepacket (constant) transverse total width.

\h By contrast, putting $\rho=0$ into eq.(23), we end up with the
expression [still with $\al \gg 1$]:

\bb |\Psi(\rho=0,\zeta)|^2 \, \simeq \, {\Ncal}^2 \; 4 \; \dis{{ {q^2 \hbar^2} \over {
q^2\hbar^2 + {1 \over {4m^2 V^2}} \zeta^2 } }} \; ,
\label{eq26}
\ee      %%%eq.26

which corresponds to

$$\Delta \zeta \ug \sqrt{e^2-1} \; {{2\al \hbar} \over {mV}} \ .$$

Solution (26) is represented in Fig.2.

\

\begin{figure}[!h]
\begin{center}
 \scalebox{.55}{\includegraphics{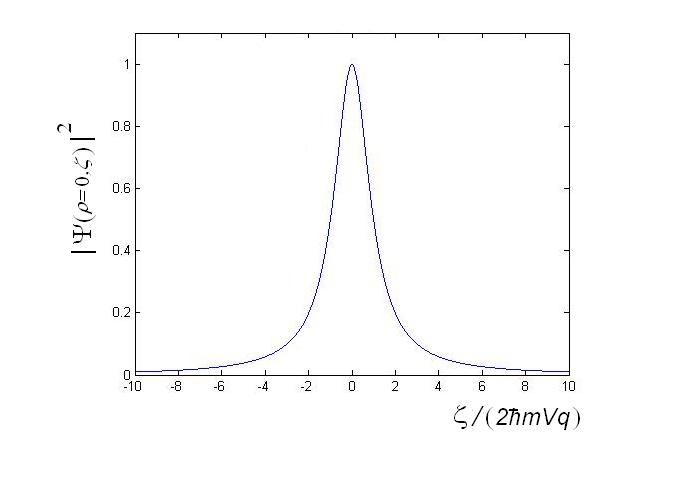}}
\end{center}
\caption{Behavior of $|\Psi(\rho=0,\zeta)|^2$ in eq.(26), as a
function of $\zeta / (2 \hbar q m V)$.} \label{fig2}
\end{figure}

\

\h Let us briefly consider a few further possible spectra. \ We shall go
on confining ourselves to the simple case of cylindrical symmetry, but
analogous solutions can be {\em easily} found also for more general
non-symmetrical cases.

\

\h {\em As the second option,} let us choose the new spectrum

\bb S(p_\rho) \ug \dis{{1 \over p_\rho} \, \dis{\e^{-q
p_\rho^2}}} \; ,
\label{eq27}
\ee      %%%eq.27

quantity $q$ being defined in eq.(22a), and condition (22b) being enforced, so
that $q \gg 1/(m^2 V^2)$ and, again, $\Delta p_\rho \ll mV$. \ Equation (20)
yields the new solution

\bb
\Psi(\rho,\zeta)\, \simeq \, \Ncal \ {1 \over 2} \; \gamma \left( 0, {\rho^2 \over
{4Q}} \right) \
\exp{[{{i2mV} \over \hbar} \, \zeta]} \; ,
\label{eq28}
\ee      %%%eq.28

where function $Q$ is defined in eq.(24), and $\gamma$, here, is
the ``incomplete gamma function".[22]

$$\gamma(0,\Acal) \ug -\gamma(-1,\Acal) - \Acal^{-1} \, \e^{-\Acal}$$

with

\hfill{$
\begin{array}{clr}
\gamma(-1,\Acal) & \equiv -\Acal^{-1} \, \e^{-\Acal} \; \Phi(1,0;\Acal)\\

             & \equiv -\Acal^{-1} \, \e^{-\Acal} [1-\Phi(1,0;\Acal)] \; ,
\end{array}
$\hfill}

\

function $\Phi$ being the ``Probability Integral", that in the present case
can be defined as

\

\hfill{$ \Phi(1,0;\Acal) \, \equiv \, \dis{{1 \over \Gamma(1)} \;
\int_0^\infty \drm x \ {{\alpha-\e^{-\Acal x}} \over {1-\e^{-x}}}} \; .
$\hfill}

\

The maximum, also for solution (27), occurs at $\rho = \zeta = 0$.

\

\h {\em As a third option,} we choose

\bb S(p_\rho) \ug \dis{{q p_\rho} \, \e^{-q p_\rho^2} \,
I_0({{s p_\rho} \over \hbar})} \,
\label{eq29}
\ee      %%%eq.29

always with $\al \gg 1$, quantity $q$ being given by eq.(22a), $s$ a
constant with the dimensions of a Length (regulating the spectrum
bandwidth), and $I_0$ being
the Modified Bessel Function; \  one gets from eq.(20) the further
new solution

\bb \Psi(\rho,\zeta) \, \simeq \, \Ncal \ {{q \hbar} \over {2Q}} \, \e^{{{i2mV}
\over \hbar} \, \zeta} \, \exp{\left[ {{s^2-\rho^2} \over {4Q}}
\right]} \; J_0 \left( {{s \rho} \over {2Q}} \right) \; .
\label{eq30}
\ee   %%%eq.30

\

\h {\em As the last option,} let us choose

\bb  S(p_\rho) \ug \dis {q p_\rho \, \e^{-q p_\rho^2} \, J_0(s p_\rho)} \; ,
\label{eq31}
\ee      %%%eq.31

from eq.(20) it follows the fourth solution

\bb
\Psi(\rho,\zeta) \simeq \Ncal \ {q \over {2Q}} \, \e^{{{i2mV} \over \hbar} \, \zeta}
\, \exp{\left[-{{s^2+\rho^2} \over {4Q}} \right]} \
I_0 \left({{s \rho} \over {2Q}} \right) \; .
\label{eq32}
\ee      %%%eq.32

\

\section{Exact Localized Solutions to the Schroedinger equation (for
arbitrary frequency spectra)}

Our aim is now to construct new analytical solutions to the Schroedinger
equation, by following an exact (not approximate) approach. \ Let us,
then, go back to eq.(1), and to its Bessel-beam solution (11), where,
as before, relation (12b) holds: \ $p_\rho = \sqrt{2mE - p_z^2}$, with
$E=\omega\hbar$.

\begin{figure}[!h]
\begin{center}
 \scalebox{.5}{\includegraphics{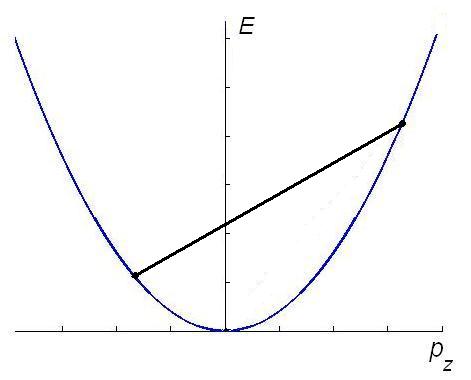}}
\end{center}
\caption{This time, the parabola and the chosen straight-line have
equations $E = p_z^2/(2m)$ and $E = Vp_z + b$, respectively. The
intersections of this straight-line with the parabola are now two,
whose corresponding values are given in eq.(33b). \ Inside the
parabola $p_\rho^2 \geq 0$.}
\label{fig3}
\end{figure}

\h The {\em condition} for obtaining a Localized Solution (cf. Fig.3) is
that

\

\hfill{$ E \ug Vp_z + b \; ,
$\hfill} (33a)          %%%eq.(33a)

\

with $b$ a positive constant (bearing the dimensions of an Energy,
and regulating the position of the chosen straight-line in the
plane $(E, p_z)$); which corresponds in particular, on using
eq.(12b), to the adoption of the integration limits

\

\hfill{$ E_{\pm} \ug mV^2 \left( 1 \pm \sqrt{1+{{2b} \over
{mV^2}}} \right) + b \; .
$\hfill} (33b)         %%%eq.33b
\setcounter{equation}{33}

\

\h Localized Solutions can therefore be obtained by the following
superpositions (integrations over the frequency, or the energy)
of Bessel-beam solutions:

\bb \Psi(\rho,z,\zeta) \ug \dis{\e^{{{-i b} \over \hbar V} z}} \;
\int_{E_{-}}^{E_{+}} \drm E \; J_0(\rho p_\rho/\hbar) \; S(E) \;
\dis{\e^{i{E \over \hbar V} \zeta}} \; ,
\label{eq34}
\ee      %%%eq.34

together with

\bb
p_\rho \ug {1 \over V} \, \sqrt {-E^2 + (2mV^2+ 2 b) E - b^2} \; .
\label{eq35}
\ee      %%%eq.35

Notice that the in eq.(34) [as well as in eq.(39) below], the solution $\Psi$
depends on $z$, besides via $\zeta$, only via a phase factor; the modulus
$|\Psi|$ of $\Psi$ goes on depending on $z$ (and on $t$) only through the
variable $\zeta \equiv z-Vt$.

\

\subsection{Particular exact Localized Solutions}

We want now to re-write the integral ${\cal{I}}$ appearing
in the r.h.s. of eq.(34) so that its integration limits are $-1$
and $+1$, respectively; that is, in the form

\

\hfill{$ {\cal{I}} \ug \dis{\int_{-1}^{1} \drm u \; S(u) \; J_0({{\rho \sqrt{P}} \over
\hbar} \sqrt{1-u^2}) \: \e^{i \, f(\zeta) \, u}} \; , $\hfill}

\

quantity $f(\zeta)$ being an arbitrary dimensionless function. To obtain this, we have to
{\em look for} a transformation of variables [with $A$ and $B$ constants, with the dimensions
of an Energy, to be determined]

\bb
E \ug Au + B
\label{eq36}
\ee      %%%eq.36

{\em such that}

\

\hfill{$
p_\rho^2 \ug P(1-u^2) \; ;  \ \ \ \ \ u_{+} = 1 \; ; \ \ \ \ \
u_{-} = -1 \; ,
$\hfill} (36')          %%%eq.(36')

\

$P$ being a suitable constant (with the dimensions of an Impulse square). \ On writing \
$V^2 p_\rho^2 \ug E \, (\hbar V^2 M - E) - b^2$, \ with $\hbar M \equiv 2m + 2 b / {V^2}$, \
after some algebra one finds that it must be

\bb
A \ug \sqrt{P} \; V \; ; \ \ \ \ \ B \ug mV^2 + b\; ; \ \ \ \ \
P \ug m^2 V^2 + 2m b \; .
\label{eq37}
\ee      %%%eqs.37

Indeed, one can verify (by some more algebra) that eqs.(36)-(37)
imply, as desired, that \ $u_{-} \ug -1$ \ and \ $u_{+} \ug 1$.

In conclusion, the transformation

\bb
E \ug mV^2 \sqrt{1+{{2 b} \over {mV^2}}} \ \; u + mV^2 + b
\label{eq39}
\ee      %%%eq.38

does actually allow writing solution (34) in the form [recall that $E=Au+B
\Longrightarrow \drm E=A\drm u$]

\bb \Psi(\rho,\eta,\zeta) \ug  \Ncal \ A \; \dis{ \e^{ {{imV} \over
\hbar} \eta} } \; \int_{-1}^{1} \drm u \; S(u) \; J_0({\rho
\over \hbar} \sqrt{P} \sqrt{1-u^2}) \; \dis{  \e^{ {{iA\zeta}
\over {\hbar V}} u } } \; , \label{eq40}
\ee      %%%eq.39

\

with

$$\eta \equiv z-vt \; ,$$

where $v \equiv V + b / (mV)$. \ Equation (39) is
exactly, analytically integrable when $S$ is a constant
or a suitable exponential.

\h Let us choose the complex exponential function
(which will easily enter as an element in a Fourier expansion)

\bb {\overline{S}(E)} \ug  a_n \; \dis{ \e^{{{2\pi i} \over D} n E} } \; ,
\label{eq41}
\ee      %%%eq.40

with $n$ an integer, and $D \equiv E_{+} - E_{-} = 2mV^2 \;
\sqrt{1+2 b / (mV^2)}$, while $a_n$ are constant quantities (with dimensions
of the inverse of an Energy). \ On remembering that $E=Au+B$,
such a spectrum can be written in terms of $u$ as

\

\hfill{$ {\ove S}(u) \ug  a_n \; \dis{\e^{i\pi n u} \; \e^{i{{2\pi} \over D} n B}}
$\hfill} (40')          %%%eq.(40')

\

(still with the dimensions of an inverse Energy). \ After some more algebra,
the analytic exact solution to the
Schroedinger equation, corresponding to spectrum (41), results to be[11]

\bb \Psi(\rho,\eta,\zeta) \ug  \Ncal a_n \; {2A} \; {\sin {Z} \over
Z} \; \dis{ \e^{{{imV} \over \hbar} \eta} \; \e^{i{{2\pi} \over D} n B}} \;  \; ,
\label{eq42}
\ee      %%%eq.41

where $A, \ B, \ P$ are given by eqs.(37) and

\bb
Z \, \equiv \, \sqrt{ \left( {A \over {\hbar V}} \zeta + n\pi \right)^2 +
{P \over \hbar^2} \rho^2} \; .
\label{eq43}
\ee      %%%eq.42

\begin{figure}[!h]
\begin{center}
 \scalebox{.5}{\includegraphics{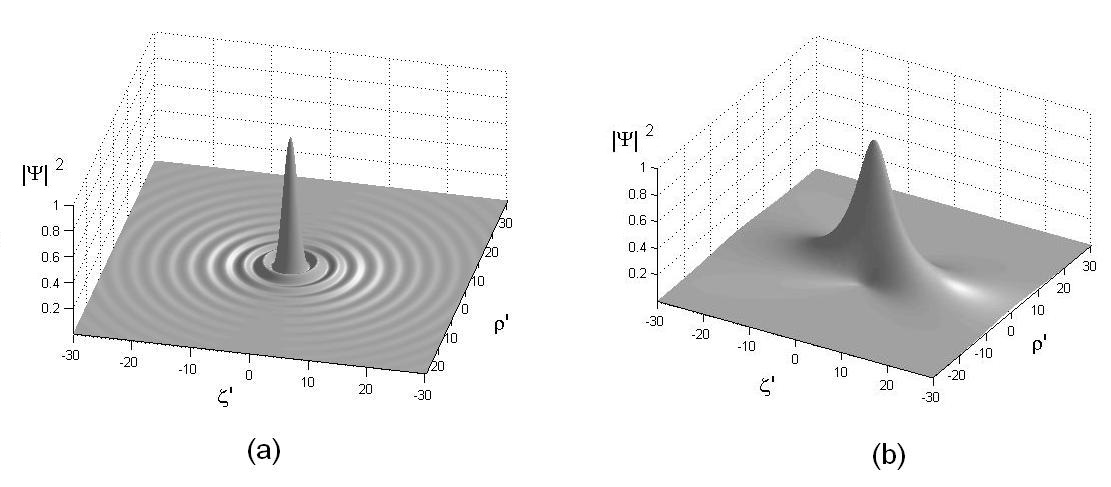}}
\end{center}
\caption{In these figures we depict an elementary solution:
namely, the square magnitude of the simplified solution, eq.(44a),
corresponding to the {\em real} spectrum $S(u) = s_0 \exp[(E -
E_+) a]$, as a function of $\rho ' \equiv \rho \sqrt{P}/\hbar$ and
of $\zeta ' \equiv \zeta \sqrt{P}/\hbar$. Quantity $a$ is a
positive number [when $a=0$ one ends up with a solutions similar
to Mckinnon's[23]], while $b$ for simplicity has be chosen equal
to zero. \ Figure (a) corresponds to $a=E_+ / 5$, while figure (b)
corresponds to $a=5 E_+$.  \ For the properties of the spectral
function (43), see the text.} \label{figs4}
\end{figure}

\h Equation (41), as we have just seen, is a particular exact Localized
Solution to the Schroedinger equation; but we are going to utilize it
essentially as {\em an element} of suitable superpositions. \ Before going on,
however, we wish to depict in Figs.\ref{figs4} an elementary solution: namely,
the square magnitude of the simple solution corresponding, in eq.(34), to the {\em real}
exponential

\bb S(E) \ug s_0 \; \dis{ \exp [a (E - E_+)] } \; ,
\label{eq44}
\ee      %%%eq.43

%\bb S(u) \ug \dis{ \exp [(u - u_+) \, g] } \; ,
%\label{eq44}
%\ee      %%%eq.43

$a$ being a positive number, endowed with the dimensions of an inverse Energy,
as well as $s_0$. \ When $a=0$, one ends up with a
solutions similar to Mckinnon's[23]. \ Spectrum (43) is
exponentially concentrated in the proximity of $E_+$, where it
reaches its maximum value; and becomes more and more concentrated
(on the left of $E_+$, of course) as the arbitrarily chosen value
of $a$ increases.  \ To perform the integration in eq.(34), it is once
more useful to operate the variable transformation (36) and go on to
eq.(39), spectrum (43) assuming now the form

$$ S(u) \ug s_0 \; e^{-a E_+} \; e^{aB} \; e^{aAu} \; .$$

Performing the integration in eq.(39), by a process similar to the one
which led us to eq.(41), in the present case we get

\

\hfill{$ \Psi(\rho,\eta,\zeta) \ug  \Ncal s_0 2V \sqrt{P} \; \dis{ \exp [i {{mV}
\over {\hbar}} \eta] \ \exp [-a V \sqrt{P}] \ {{\sin Y} \over Y} }
$\hfill} (44a)          %%%eq.(44a)

\

where

\

\hfill{$ Y \; \equiv \; {{\sqrt{P}} \over \hbar} \, \sqrt{\rho^2 -
(\hbar aV+i\zeta)^2} \; ,
$\hfill} (44b)         %%%eq.44b
\setcounter{equation}{44}

\

quantity $P$ having been defined in eq.(37); and one should remember that
$\eta \equiv z-vt$ is a function of $b$.

\h Equations (44) appear to be the simplest closed-form solutions
(see Figs.\ref{figs4}) to the Schroedinger equation, since they do
not need any recourse to series expansions of the type exploited
in the following Subsection. However, the solutions that we shell
construct below can correspond to spectra more general than (43);
for instance, to the gaussian spectrum, which possesses two
advantage w.r.t. spectrum (43): it can be easily centered around
any value of $u$, that is, around any value $\bar{E}$ of $E$ in
the interval [$E_-, E_+$], and, when increasing its concentration
in the surrounding of $\bar{E}$, its ``spot" transverse width does
not increase, at variance with what happens for spectrum (43).
Anyway, the exact solutions (44) are noticeable, since they are
really the simplest ones.

\h Some physical (interesting) comments on the results in eqs.(44)
and Figs.4 will appear elsewhere.  Here, let us add only a few
further Figures and some brief comments.  Let us first recall
that, as predicted in the first one of Refs.[6], the Localized
(Nondiffracting) Solutions to the ordinary wave equations resulted
to be roughly {\em ball-like} when their peak-velocity is
subluminal[11], and {\em X-shaped\/}[4,7] when superluminal.

Now, normalizing $\rho$ and $\zeta$, we can write eq.(44b) as

$$Y \ug \sqrt{{\rho'}^2 - (\Aove + i\zeta')^2}$$

with \ $\rho' \equiv \sqrt{P} \rho / \hbar$ \ and \ $\zeta' \equiv
\sqrt{P} \zeta / \hbar$, \ quantity $P$ being given by the last
one of eqs.(37), namely $P = m^2 V^2 +2mb$, \ while \ $\Aove
\equiv aA = \sqrt{P} a V$. \ For simplicity, let us confine
ourselves to the case $b=0$, forgetting now about the more
interesting cases with $b \neq 0$; \ therefore, it will hold the
simple relation

$$\Aove = maV^2 \ .$$

In the present case of the Schroedinger equation, we can observe
the following.

\h If we choose $\Aove = 0$, which can be associated with $V=0$, we get the solutions in
Figs.5: that is, a ball-like structure.

\begin{figure}[!h]
\begin{center}
 \scalebox{0.6}{\includegraphics{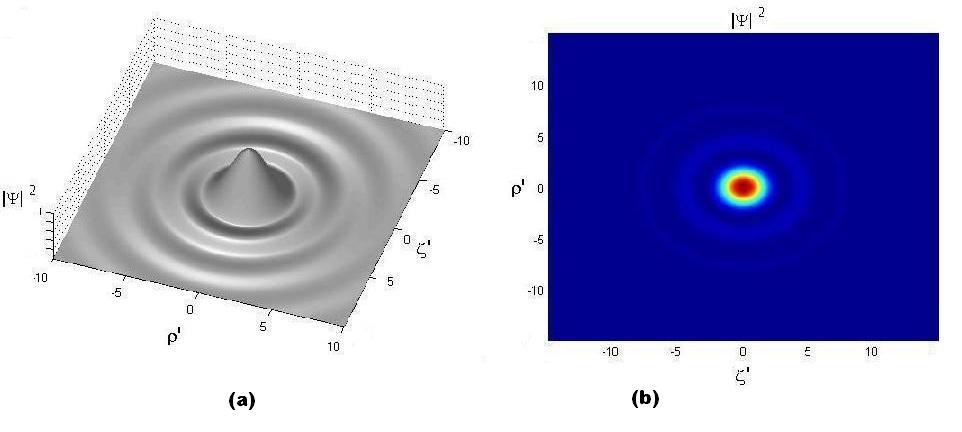}}
\end{center}
\caption{In these, and the following Figures 6, 7 and 8, we depict the square magnitude
of some more solutions of the type (44a), normalized with respect to $\rho$ and $\zeta$;
still assuming for simplicity $b=0$, so that $\Aove = maV^2$.  The present figures show
the ``ball-like"
structure that one gets, as expected, when $\Aove = 0$ (see the text, also for the
definitions of $\rho'$ and $\zeta'$). \ Fig.(b) shows the projection on the plane
($\rho', \ \zeta'$) of the 3D plot shown in Fig.(a).}
\label{figs5}
\end{figure}

\

\

\h By contrast, if we increase the value of $\Aove$, by choosing e.g. $\Aove=20$ (which can be
associated with larger speeds), one notices that also a X-shaped structure starts to
contribute: See, e.g., Fig.6.

\begin{figure}[!h]
\begin{center}
 \scalebox{1.8}{\includegraphics{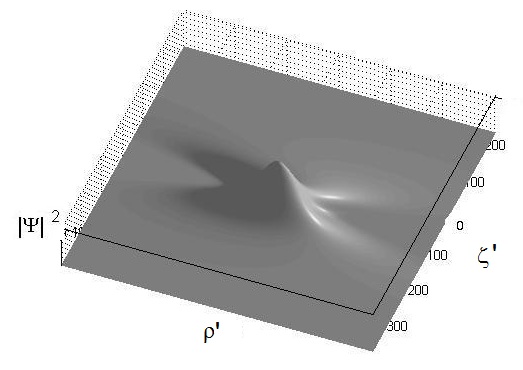}}
\end{center}
\caption{The solution, under all the previous conditions, with an increased value of
$\Aove$, namely with $\Aove = 20$. An X-shaped structure starts to appear, contributing
to the general form of the solution (see the text).}
\label{fig6}
\end{figure}

\

\

\h To have a preliminary idea of the ``internal structure" of our soliton-like solutions
to the (ordinary) Schroedinger equation, let us plot, instead of the square magnitude
of $\Psi$, its real or imaginary part: Let us choose its real part, or rather the
square of its real part.  \ Then even in the $\Aove = 0$ case one starts to
see the appearance of the X shape, which becomes more and more evident as the value of
$\Aove$ increases: In Figs.7  we show the
projections on the plane ($\zeta', \ \rho'$) of the real-part square for the solutions
with $\Aove = 5$  and $\Aove = 50$, respectively. \ Further attention to such aspects
will be paid elsewhere.

\begin{figure}[!h]
\begin{center}
 \scalebox{1.8}{\includegraphics{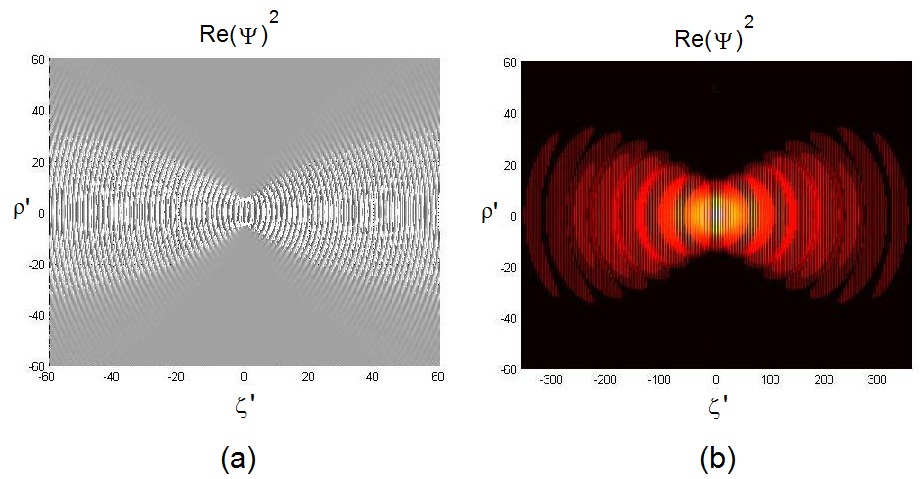}}
\end{center}
\caption{To get a preliminary idea of the ``internal structure" of our soliton-like
solutions, it is useful to have recourse (see the text) to the real part of $\Psi$.
In these Figures we plot the projections on the plane ($\zeta', \ \rho'$) of the
real-part square for the solutions  with  $\Aove = 5$  (figure (a)) and $\Aove = 50$
(figure (b)), respectively.}
\label{figs7}
\end{figure}

\

\

\h But the (square of the) real part of $\Psi$ does show, in 3D, also some ``internal oscillations":
Cf., e.g., Fig.8 corresponding to the value $\Aove = 5$. \ We
shall face elsewhere, however, topics like their possible connections with the de Broglie
picture of quantum particles, et alia.

\begin{figure}[!h]
\begin{center}
 \scalebox{2.3}{\includegraphics{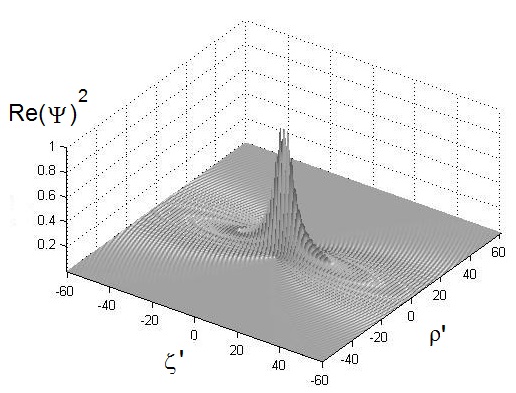}}
\end{center}
\caption{The (square of the) real part of $\Psi$ shows, in 3D, also some ``internal
oscillations": this Figure corresponds, e.g., to the value $\Aove = 5$.}
\label{fig8}
\end{figure}

\

\

\

\subsection{A general exact Localized Solution}

Let us go back to our spectrum $S(E)$ in eq.(40).  Since in our fundamental
equation (34) the integration interval is limited [$E_- < E < E_+$], in such
an interval {\em any} spectral function $S(E)$ whatever can be expanded into
the Fourier series

\bb S(E) \ug \sum_{n=-\infty}^{\infty} \, a_n \, \dis{
\e^{i{{2\pi} \over D} n E} } \; ,
\label{eq46}
\ee      %%%eq.45

with

\bb a_n \ug {1 \over D} \dis{ \int_{E_-}^{E_+} \drm E \; S(E) \;
\e^{-i{{2\pi} \over D} n E} } \; ,
\label{eq47}
\ee      %%%eq.46

quantity $S(E)$ being an {\em arbitrary} function, and $D$ being still defined as
$D \equiv E_{+} - E_{-}$.

\h Inserting eq.(45) into eq.(34), and following the same
procedure exploited in the previous Subsection (in particular,
going on again from $E$ to the new variable $u$), we end up ---after
normalization--- with the {\em
general exact localized solution} to the Schroedinger equation:

\bb \Psi(\rho,\eta,\zeta) \ug \Ncal \, {2A} \, \dis{ e^{i
{{mV} \over {\hbar}}\eta}}  \
\sum_{n=-\infty}^{\infty} \, a_n \; \dis{ \exp{[{i{{2\pi} \over D}
n B}]} } \; {\sin {Z} \over Z} \; , \label{eq48}
\ee      %%%eq.47

where $Z$ is defined in eq.(42), and the coefficients $a_n$ are
given by eq.(46).

\h It is worthwhile to note that, even when truncating the series
in eq.(47) at a certain value $n=N$, the solutions obtained is
{\em still} an exact LS of the Schroedinger equation!

\

\

\section{About finite-energy Localized Solutions to the Schroedinger equation}

The solutions found above, even if very instructive, are ideal solutions which are not square
integrable; and cannot be accepted in QM. It is important, therefore, to show how to
construct finite-energy solutions.

\h Let us obtain localized solution to the Schroedinger equation
endowed with {\em finite energy,} by starting from eqs.(44). First
of all, one has to integrate over $b$ by adopting a spectrum
$S(b)$ strongly bumped around a value $b_0$: We already know,
indeed, that spectra of this type are required in order to get
solutions that are non-diffracting all along a certain
field-depth.

\h Then, it can be easily seen that the finite-energy solution,
$\Psi_{\rm fe}$, can be preliminarily written as

\bb \Psi_{\rm fe} \ug \Ncal \ {{s_0 V \sqrt{P}} \over {iY}} \ (I_- -
I_+) \; , \label{eq49}
\ee      %%%eq.48

where $I_-$ and $I_+$ are two (dimensionless) integrations over
$b$ from 0 to infinity (quantity $b$ having been defined in
eq.(33a), and therefore having the dimensions of an Energy), while
$s_0$ appears in eq.(43).

Let us now pass from $b$, defined in eq.(33a), to the new variable $w \equiv \sqrt{P}$.
One has to choose a spectrum $S(w)$ corresponding to a $S(b)$
concentrated around a specific value of $b$; let us therefore adopt
the gaussian function

\bb  \Scal (w) \ug   {{ m \sqrt{q}} \over {\sqrt{\pi} \hbar w}} \;
\exp [-q(w - w_0)^2] \; , \label{eq50}
\ee      %%%eq.49

with $w_0 > mV > 0$.

When we go on from $b$ to the new variable $w \equiv \sqrt{P}$ (where $P$ depends on $b$),
the two quantities $I_-$ and $I_+$ become integrations over $w$ from
$mV$ to $\infty$.  After further calculations, and using relation 3.322.1 in ref.[22],
one obtains that

\bb I_{\pm} \ug {{\sqrt{q}}\over {U}} \, e^{-qw_0} \,
e^{{{imV} \over 2\hbar} z} \,  \exp {[{W_{\pm}^2 \over U^2}]} \
\left[ 1 - \Phi \left( {W_{\pm} \over U} + {{mV} \over {2}} U
\right) \right] \; , \label{eq51}
\ee      %%%eq.50

where

\

$$ U \equiv 2 \sqrt{q+{{i\hbar} \over {2m}} t} \, ; \ \ \ \ \ W_{\pm} \equiv -2qw_0 +
aV \pm i{Y \over \sqrt{P}} \; , $$

\

quantity $Y$ having been defined in eq.(44b).

\h We have therefore shown that realistic (finite-energy)
Localized Solutions exist also to the Schroedinger equation; they
will be non-diffracting only till a certain finite distance (depth
of field).  The analysis of explicit, particular examples will be
presented elsewhere.

\

\

\section{The case of non-free particles}

Let us consider now the case of a particle in the presence of a
potential: for simplicity, let us confine ourselves to the case of
a {\em cylindrical potential.}

\

\h Namely, let us consider the Schroedinger equation with a potential of
the type $U(\rho)$:

\bb -\frac{\hbar^2}{2m}\left(\nabla^2_{\bot} + \frac{\pa^2}{\pa
z^2} \right)\psi  + U(\rho)\psi  -i\hbar \frac{\pa \psi}{\pa t}
\ug 0 \label{sch} \ee         %%%eq.51

\h Now, we can use the method of separation of variables writing
$\psi = R(x,y)Z(z)T(t)$. With this, we get the well known
solutions

\bb T = \dis{ e^{-\frac{i}{\hbar}E\,t} }  \ee       %%%eq.52

\

\bb Z = \dis { e^{i p_z z / \hbar} }  \ee          %%%eq.53

and the eigenvalue equation

\bb - \hbar^2 \, \nabla^2_{\bot}R + 2m \, U(\rho) \, R \ug \Lambda^2 R
\label{eqR} \ee              %%%eq.54

with

\bb \La^2 \ug 2m E - p_z^2 \label{la} \ee         %%%eq.55

\h Supposing a potential $U(\rho)$ that only allows transverse
bound states (as the parabolic potential), we will find
eigenfunctions $R_n(x,y)$ and discrete (degenerate) eigenvalues
$\La_n^2$.

\h We can construct more general solutions

\bb \Psi = \sum_{n} f_n R_n(x,y)e^{ik_z z / \hbar}e^{-\frac{i}{\hbar}E t}
\label{geral} \ee        %%%eq.56

with

\bb 2m E \ug  p_z^2 + \La^2_n \label{lan} \ee   %%%eq.57

\h Considering $p_z \geq 0$ (forward propagation), the constraint
(\ref{lan}) defines a set of parabolas (something like the modes
in a waveguide: Cf. Refs.16). Chosen a certain  $\La_n^2$, once a value for $p_z$ is
given, the value of $E$ gets fixed.

\h To obtain from (\ref{geral}) a train of localized pulses, i.e.,  a wavefunction
$\Psi(x,y,z-Vt)$, we must have

\bb E \ug V p_z  \label{cond} \ee        %%%eq.58

\h So, from conditions (\ref{lan}) and (\ref{cond}), $p_z$ must
assume the values

\bb p_z \ug mV \left(1 \pm \sqrt{1 -
\frac{1}{m^2V^2}\La_n^2} \right) \label{kz} \ee      %%%eq.59

with

\bb \La_n \leq mV \label{condlan} \ee     %%%eq.60

\h Figure 9 illustrates the situation. The values to $E$ and $p_z$
that furnish localized pulse trains are given by the intersection
between the parabolas defined by eq.(\ref{lan}) and the straight
line defined by eqs.(\ref{cond}). Note that in these cases the
series (\ref{geral}) will be always truncated (finite number of
terms), due the condition (\ref{condlan}). We also have to note
that, for any given $\la_n^2$, one gets two possible values of
$k_z$ (see eq.(\ref{kz})), as it can be observed from Fig.9, in
which the straight line cuts each parabola twice.

%%\newpage

\begin{figure}[!h]
\begin{center}
 \scalebox{.6}{\includegraphics{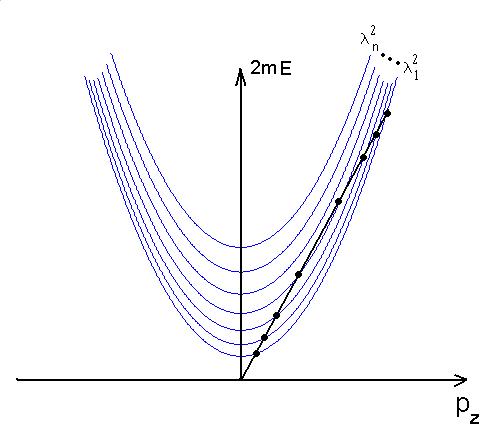}}
\end{center}
\caption{In the case of a particle in the presence of a
cylindrical potential, the values to $E$ and $p_z$ that furnish
Localized Pulse trains are given by the intersection between the
parabolas in eq.(\ref{lan}) and the straight line in
eq.(\ref{cond}): see the text. It can be noticed that, for any
given $\la_n^2$, one gets two possible values of $k_z$ (cf.
eq.(\ref{kz})), since the straight line cuts each parabola twice.
\ See the text, and cf. also Refs.[17].} \label{fig9}
\end{figure}                 %%%Fig.9

\

\h For our purpose, the superposition has to be

\bb \Psi(x,y,z-Vt) \ug \sum_{n} f_n R_n(x,y)e^{ip_{zn}(z-Vt) / \hbar}
\label{geral2} \ee     %%%eq.61

with

\bb p_z \ug mV \left(1 \pm \sqrt{1 -
\frac{1}{m^2V^2}\La_n^2} \right) \ee        %%%eq.62

  and

\bb \La_n \leq mV \ee        %%%eq.63

\h In principle, any set of coefficients $f_n$ will furnish {\em trains
of localized waves}.

\h Observation1: If we look for a square-integrable wave function,
we can start from superposition (\ref{geral}) and integrate its
terms over $p_{z}$ around each $p_{zn}$, respectively (as we
already did in our papers on X-type pulses propagating along
wave-guides[17]). But in the present case, in general, the
group-velocities defined at the points $p_{zn}$ will {\em not} be
the same, as it happened in the waveguide case; and we will
therefore meet a kind of intermodal dispersion, besides the
group-velocity dispersion. Let us recall, incidentally, that such
an intermodal dispersion did not occur in the case of X-type
waves, traveling in metallic wave-guides, due the peculiar fact
that the group-velocities defined at those points were always the
same ). After the integration, we can obtain an envelope with a
train of pulses (or just one pulse) inside it. The envelope will
suffer dispersion, but the train of pulses inside it will not.

\h More general localized wave trains can be
obtained using the relation $E \ug V p_z + b$, with $b$
a positive constant.

\h In the case of potentials like $U(\rho)$, one can search for
solutions with cylindrical symmetry, for simplicity. However,
solutions without this symmetry can be investigated: and they will
be interesting for an analysis of angular momentum.

\section{Acknowledgments}

The authors are grateful to Claudio Conti, Hugo E. Hern\'andez-Figueroa e Peeter
Saari for many stimulating contacts and discussions. After the completion of this
work (see, e.g., our e-print arXiv:1008.3087[quant-ph]), we came to know that some
work on the same topic, by following different paths, has been done also by
I.B.Besieris and A.M.Shaarawi (``Localized traveling wave solutions to the 3D
Schroedinger equation": unpublished): And we are grateful to I.M.Besieris for
such a piece of information.

\newpage

%\section{Bibliography}

\end{document}